\documentclass[pra,twocolumn,preprintnumbers,amsmath,amssymb]{revtex4}
\usepackage{graphics}
\usepackage{psfrag}
\usepackage{hyperref}

\begin{document}

\title{Superfluid Bose gas in two dimensions}
\author{S. Floerchinger}
\author{C. Wetterich}
\affiliation{Institut f\"{u}r Theoretische Physik\\Universit\"at Heidelberg\\Philosophenweg 16, D-69120 Heidelberg}

\begin{abstract}
We investigate Bose-Einstein condensation for ultracold bosonic atoms in two-dimensional systems. The functional renormalization group for the average action allows us to follow the effective interactions from molecular scales (microphysics) to the characteristic extension of the probe $l$ (macrophysics). In two dimensions the scale dependence of the dimensionless interaction strength $\lambda$ is logarithmic. Furthermore, for large $l$ the frequency dependence of the inverse propagator becomes quadratic. We find an upper bound for $\lambda$, and for large $\lambda$ substantial deviations from the Bogoliubov results for the condensate depletion, the dispersion relation and the sound velocity. The melting of the condensate above the critical temperature $T_c$ is associated to a phase transition of the Kosterlitz-Thouless type. The critical temperature in units of the density, $T_c/n$, vanishes for $l\to\infty$ logarithmically.
\end{abstract}

\pacs{}

\maketitle

\section{Introduction}
\label{sectIntroduction}
Bose-Einstein condensation and superfluidity for cold nonrelativistic atoms can be experimentally investigated in systems of various dimensions \cite{ExperimentBECReview}. Two dimensional systems can be achieved by building asymmetric traps, resulting in different characteristic sizes for one ``transverse extension'' $l_T$ and two ``longitudinal extensions'' $l$  of the atom cloud \cite{Experiment2D}. For $l \gg l_T$ the system behaves effectively two-dimensional for all modes with momenta $\vec{q}^2\lesssim l_T^{-2}$. From the two-dimensional point of view, $l_T$ sets the length scale for microphysics -- it may be as small as a characteristic molecular scale. On the other hand, the effective size of the probe $l$ sets the scale for macrophysics, in particular for the thermodynamic observables.

Two-dimensional superfluidity shows particular features. In the vacuum, the interaction strength $\lambda$ is dimensionless such that the scale dependence of $\lambda$ is logarithmic \cite{Lapidus}. The Bogoliubov theory with a fixed small $\lambda$ predicts at zero temperature a divergence of the occupation numbers for small $q=|\vec{q}|$, $n(\vec{q})\sim n_C\, \delta^{(2)}(\vec{q})$ \cite{Bogoliubov}. In the infinite volume limit, a nonvanishing condensate $n_c=\bar{\rho}_0$ is allowed only for $T=0$, while it must vanish for $T>0$ due to the Mermin-Wagner theorem \cite{MerminWagnerTheorem}.  On the other hand, one expects a critical temperature $T_c$ where the superfluid density $\rho_0$ jumps by a finite amount according to the behavior for a Kosterlitz-Thouless phase transition \cite{KosterlitzThouless}. We will see that $T_c/n$ (with $n$ the atom-density) vanishes in the infinite volume limit $l\to \infty$. Experimentally, however, a Bose-Einstein condensate can be observed for temperatures below a nonvanishing critical temperature $T_c$ -- at first sight in contradiction to the theoretical predictions for the infinite volume limit.

A resolution of these puzzles is related to the simple observation that for all practical purposes the macroscopic size $l$ remains finite. Typically, there will be a dependence of the characteristic dimensionless quantities as $\bar{\rho}_0/n$, $T_c/n$ or $\lambda$ on the scale $l$. This dependence is only logarithmic. While $\lambda(n=T=0, l\to \infty)=0$, $(\bar{\rho}_0/n)(T\neq0, l\to \infty)=0$, $(T_c/n)(l\to0)=0$, in accordance with general theorems, even a large finite $l$ still leads to nonzero values of these quantities, as observed in experiment. 

An appropriate tool for studying the scale dependence of various quantities is the functional renormalization for the average action \cite{Wetterich:1992yh}. Within a two-dimensional context, this starts with a given microphysical or classical action at some ultraviolet momentum scale $\Lambda_\text{UV}\sim l_T^{-1}$. Then a scale dependent effective action $\Gamma_k$ is defined by integrating out successively the modes with momenta $|\vec{q}|$ larger than some infrared cutoff scale $k$. The flow of $\Gamma_k$ with $k$ obeys an exact renormalization group equation, which we solve approximately by truncation of the most general form of $\Gamma_k$. When $k$ reaches a scale $k_\text{ph}\sim l^{-1}$, all fluctuations are included since no larger wavelength are present in a finite size system. The experimentally relevant quantities and the dependence on $l$ can be obtained from $\Gamma_{k_\text{ph}}$.  

This paper investigates the functional renormalization group for a nonrelativistic Bose gas in two spatial dimensions. A priori, this method is not restricted to a perturbative treatment for a small coupling $\lambda$ and can therefore be used to investigate the effects of fluctuations beyond the Bogoliubov approximation. We assume that the interaction can be approximated by a contact term. This should be appropriate especially for dilute gases where the details of the interaction are not relevant. More explicitly our microscopic action reads
\begin{equation}
S[\phi]=\int_x \,{\Big \{}\phi^*\,(\partial_\tau-\Delta-\mu)\,\phi\,+\,\frac{1}{2}\lambda(\phi^*\phi)^2{\Big \}}.
\label{microscopicaction}
\end{equation}
We work with the functional integral within the Matsubara formalism. The ``Euclidean time'' $\tau$ parameterizes a torus with circumference $T^{-1}$, and the bosonic fields $\phi$ are complex, depending on the two-dimensional coordinates $\vec{x}$ and $\tau$, $x=(\vec{x},\tau)$. In addition to the temperature $T$ and the chemical potential $\mu$ the action involves the dimensionless coupling $\lambda$ as a parameter. We use $\hbar=k_B=1$ and units of energy where the mass of the atoms is set to $1/2$, $2M=1$.  

All macroscopic properties following from the microscopic model \eqref{microscopicaction} can be obtained from the full effective action $\Gamma[\phi]$. For example, the thermodynamic properties can be inferred from the grand canonical partition function $Z$ or the corresponding grand potential $\Phi_G= -T \,\text{ln}\,Z$. The latter is related to the effective action by
\begin{equation}
\Gamma[\phi_\text{eq}]=\Phi_G/T,
\end{equation}
where the expectation value $\phi_\text{eq}$ is obtained as a solution of the field equation
\begin{equation}
\frac{\delta}{\delta \phi}\Gamma[\phi_\text{eq}]=0.
\end{equation}

Since the effective action is the generating functional of the one particle irreducible (1PI) correlation functions we can also derive dynamical properties from $\Gamma[\phi]$. For a homogeneous field $\phi_\text{eq}$ the full propagator $G(p)$ is obtained from
\begin{eqnarray}
\nonumber
(\Gamma^{(2)})_{ij}(p_1,p_2) &=& \frac{\delta}{\delta\phi_i^*(p_1)}\frac{\delta}{\delta \phi_j(p_2)}\Gamma[\phi]{\bigg |}_{\phi=\phi_\text{eq}}\\
&=& (G^{-1})_{ij}(p_1)\delta(p_1-p_2).
\end{eqnarray}
Here we employ a basis with two real fields, $i=1,2$, $\phi=\frac{1}{\sqrt{2}}(\phi_1+i\phi_2)$, $\phi_i^*(p)=\phi_i(-p)$. 

The calculation of the functional $\Gamma[\phi]$ is not an easy task -- it includes the summation of infinitely many diagrams in perturbation theory. Our method determines $\Gamma[\phi]$ by using an exact flow equation for the scale dependent effective action $\Gamma_k[\phi]$-- the average action. Due to the presence of an infrared cutoff at the scale $k$ only the quantum and thermal fluctuations with momenta $\vec{q}^2\gtrsim k^2$ are included. In the infinite volume limit $\Gamma_k$ interpolates between the microscopic action and the full effective action
\begin{eqnarray}
\nonumber
\Gamma_{k}[\phi] & \rightarrow S[\phi] & \quad (k\rightarrow \Lambda),\\
\Gamma_{k}[\phi] & \rightarrow \Gamma[\phi] & \quad (k\rightarrow 0).
\end{eqnarray}
For a system with finite size $l$ we are interested in $\Gamma_{k_\text{ph}}$, $k_\text{ph}=l^{-1}$. If statistical quantities for finite size systems depend only weakly on $l$, they  can be evaluated from $\Gamma_{k_\text{ph}}$ in the same way as their thermodynamic infinite volume limit follows from $\Gamma$. Details of the geometry etc. essentially concern the appropriate factor between $k_\text{ph}$ and $l^{-1}$. 

The average action obeys the exact flow equation \cite{Wetterich:1992yh, Berges:2000ew}
\begin{equation}
\partial_k\,\Gamma_k[\phi]=\frac{1}{2}\text{Tr}(\Gamma_k^{(2)}[\phi]+R_k)^{-1}\partial_k R_k.
\label{eqFlowequation}
\end{equation}
Despite its simple one loop structure it includes all orders in perturbation theory as well as nonperturbative effects. Nevertheless, in praxis this equation is not solvable analytically since it is a differential equation for a functional. Approximate solutions can be found by truncating the space of possible functionals $\Gamma_k[\phi]$ to a manageable size. Such a truncation does not have to rely on an expansion in a small parameter, as it it the case for perturbation theory. Because of this, the method is nonperturbative and also applicable for large interaction strength. 

In this work we use a very simple truncation which includes terms with up to two derivatives
\begin{eqnarray}
\nonumber
\Gamma_k&=&\int_x\bigg{\{} \bar{\phi}^*\left(\bar{S}\partial_\tau-\bar{A}\Delta-\bar{V}\partial_\tau^2\right)\bar{\phi}\\
&&+2\bar{V}(\mu-\mu_0)\,\bar{\phi}^*\left(\partial_\tau-\Delta\right)\bar{\phi}+\bar{U}(\bar{\rho},\mu)\bigg{\}}.
\label{eq:truncationbare}
\end{eqnarray}
Here the effective potential $\bar{U}$ contains no derivatives and is a function of $\bar{\rho}=\bar{\phi}^*\bar{\phi}$. At the end we will evaluate all quantities for a fixed chemical potential $\mu_0$ -- the term $\sim(\mu-\mu_0)$ will be employed for an easy evaluation of $\partial \Gamma_k / \partial \mu$, as needed for the density. In addition to the couplings parametrizing $\bar{U}$ (see below) our truncation contains three further $k$-dependent (``running'') couplings $\bar{A}$, $\bar{S}$ and $\bar{V}$. The particular form of our truncation is motivated by a more systematic derivative expansion and an analysis of symmetry constraints (Ward identities) in Ref. \cite{FW}. (This reference may also be consulted for details of the formalism and the flow equations.) A truncation similar to Eq. \eqref{eq:truncationbare} has been investigated at zero temperature in Refs. \cite{WetterichQPT2008, Dupuis}. A simpler truncation was used in Ref. \cite{Stoof}. In terms of renormalized fields $\phi=\bar{A}^{1/2}\bar{\phi}$, $\rho=\bar{A}\bar{\rho}$, renormalized kinetic coefficients $S=\frac{\bar{S}}{\bar{A}}$, $V=\frac{\bar{V}}{\bar{A}}$ and effective potential $U(\rho,\mu)=\bar{U}(\bar{\rho},\mu)$, Eq. \eqref{eq:truncationbare} reads
\begin{eqnarray}
\nonumber
\Gamma_k &=& \int_x\bigg{\{} \phi^*\left(S\partial_\tau-\Delta-V\partial_\tau^2\right)\phi\\
&&+2V(\mu-\mu_0)\, \phi^*\left(\partial_\tau-\Delta\right)\phi+U(\rho,\mu)\bigg{\}}.
\label{eqSimpleTruncation}
\end{eqnarray}

For the effective potential, we use an expansion around the $k$-dependent location of the minimum $\rho_0(k)$, and the $k$-independent value of the chemical potential $\mu_0$ that corresponds to the physical particle number density $n$. We determine $\rho_0(k)$ and $\mu_0$ by the requirements
\begin{eqnarray}
\nonumber
(\partial_\rho U)(\rho_0(k),\mu_0)&=0\quad&\text{for all}\,k\\
-(\partial_\mu U)(\rho_0,\mu_0)&=n\quad&\text{at}\,\,k=k_\text{ph}.
\end{eqnarray}
More explicitly, we take a truncation for $U(\rho,\mu)$ of the form
\begin{eqnarray}
\nonumber
U(\rho,\mu)&=&U(\rho_0,\mu_0)-n_k(\mu-\mu_0)\\
\nonumber
&&+\left[m^2+\alpha(\mu-\mu_0)\right](\rho-\rho_0)\\
&&+\frac{1}{2}\left[\lambda+\beta(\mu-\mu_0)\right](\rho-\rho_0)^2.
\end{eqnarray}
In the symmetric phase we have $\rho_0=0$, while in the phase with spontaneous symmetry breaking, we have $m^2=0$. The atom density $n=-\partial U / \partial \mu$ corresponds to $n_k$ in the limit $k\to k_\text{ph}$. 

In total, we have at fixed chemical potential $\mu=\mu_0$ and fixed temperature $T$ four running renormalized couplings $\rho_0(k)$, $\lambda(k)$, $S(k)$, and $V(k)$. (In the symmetric phase $\rho_0$ is replaced by $m^2$.) In addition, we need the anomalous dimension $\eta=-k\partial_k \text{ln} \bar{A}$. To determine the flow of the density $n_k$, we also need the dependence of the effective potential on $(\mu-\mu_0)$ and we therefore include the couplings $\alpha(k)$, $\beta(k)$. The precise procedure how the flow equation for the average action is projected to flow equations for the running couplings is explained in Ref. \cite{FW}. The flow equation for the effective potential $U_k(\rho,\mu)$ and the kinetic coefficients $S$ and $V$ as well as the anomalous dimension $\eta$ are given there for arbitrary dimension $d$. In this paper we apply these flow equations for $d=2$.

\section{Scale dependence of interaction strength}
It is well known that in two dimensions the scattering properties cannot be determined by a scattering length as it is the case in three dimensions. In experiments where a tightly confining harmonic potential restricts the dynamics of a bose gas to two dimensions, the interaction strength has a logarithmic energy dependence in the two-dimensional regime \cite{Stoof}. For low energies and in the limit of vanishing momentum the scattering amplitude vanishes. In our formalism this is reflected by the logarithmic running of the interaction strength $\lambda(k)$ in the vacuum, where both temperature and density vanish. In general, the flow equations in vacuum describe the physics of few particles like for example the scattering properties or binding energies. Following the calculation of Ref. \cite{FW}, we find the flow equations for the interaction strength ($t=\text{ln}(k/\Lambda)$)
\begin{equation}
\partial_t \lambda = \frac{\lambda^2\left( k^2 - m^2 \right)}{4k^2\,
    {\pi }\,S}\,\theta(k^2-m^2).
\label{eqflowvacuummlambda}
\end{equation}
Since in vacuum the propagator is not renormalized, $\partial_t S=\partial_t V=\partial_t \bar{A}=\partial_t m^2=0$, we set $S=1$ and $V=0$ on the right-hand side of Eq. \eqref{eqflowvacuummlambda}. The vacuum corresponds to $m^2=0$ \cite{FW} and we obtain the flow equation
\begin{equation}
\partial_t \lambda=\frac{\lambda^2}{4\pi}.
\label{eqflowoflambdainvacuum}
\end{equation}
The vacuum flow is purely driven by quantum fluctuations. It will be modified by the thermal fluctuations for $T\neq 0$ and for nonzero density $n$.

The solution of Eq. \eqref{eqflowoflambdainvacuum}
\begin{equation}
\lambda(k)=\frac{1}{\frac{1}{\lambda_\Lambda}+\frac{1}{4\pi}\text{ln}(\Lambda/k)}
\label{eqlambdakofk}
\end{equation}
goes to zero logarithmically for $k\rightarrow0$, $\lambda(k=0)= 0$. In contrast to the three-dimensional system, the flow of the interaction strength $\lambda$ does not stop in two dimensions. To relate the microscopic parameter $\lambda_\Lambda$ to experiments exploring the scattering properties, we have to choose a momentum scale $q_\text{exp}$, where experiments are performed. To a good approximation the relevant interaction strength can be computed from Eq. \eqref{eqlambdakofk} by setting $k=q_\text{exp}$. If not specified otherwise, we will use a renormalized coupling $\lambda=\lambda(k_\text{ph})$.

For our calculation we also have to use a microscopic scale $\Lambda$ below which our approximation of an effectively two dimensional theory with pointlike interaction becomes valid. Our two-dimensional computation only includes the effect of fluctuations with momenta smaller then $\Lambda$. In experiments $\Lambda^{-1}$ is usually given either by the range of the van der Waals interaction or by the length scale of the potential that confines the system to two dimensions. We choose in the following
\begin{equation}
\Lambda=10,\quad k_\text{ph}=10^{-2}. 
\end{equation}
At this stage the momentum or length units are arbitrary, but we will later often choose the density to be $n=1$, so that we measure length effectively in units of the interparticle spacing $n^{-1/2}$. For typical experiments with ultracold bosonic alkali atoms one has $n^{-1/2}\approx 10^{-4}\text{cm}$. 
 
\begin{figure}
\includegraphics{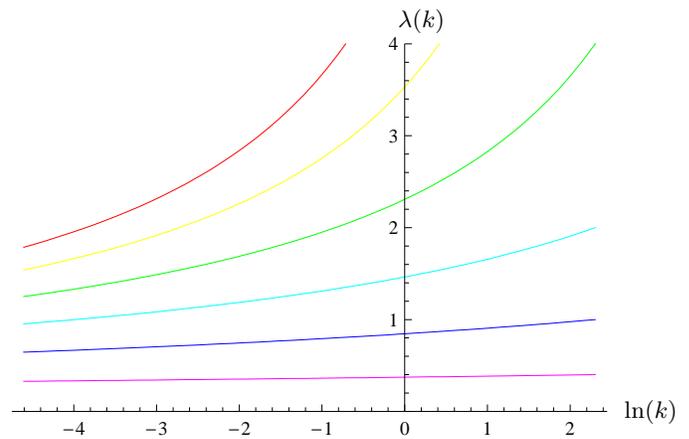}
\caption{(Color online) Flow of the interaction strength $\lambda(k)$ at zero temperature and density for different initial values $\lambda_\Lambda=100$, $\lambda_\Lambda=10$, $\lambda_\Lambda=4$, $\lambda_\Lambda=2$, $\lambda_\Lambda=1$, and $\lambda_\Lambda=0.4$ (from top to bottom).}
\label{figflowoflambda}
\end{figure}

\begin{figure}
\includegraphics{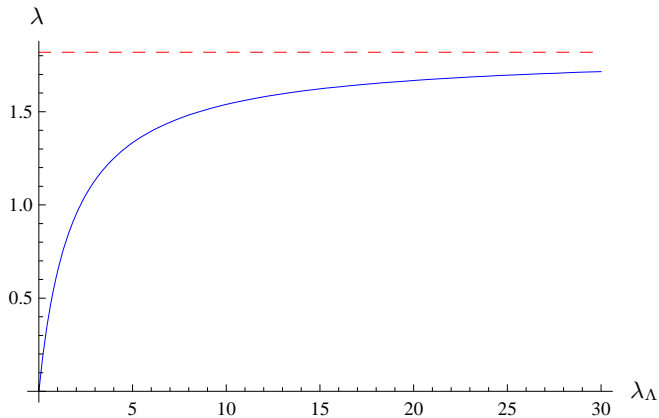}
\caption{(Color online) Interaction strength $\lambda$ at the macroscopic scale $k_\text{ph}=10^{-2}$ in dependence on the microscopic interaction strength $\lambda_\Lambda$ at $\Lambda=10$ (solid). The upper bound $\lambda_{\text{max}}=\frac{4 \pi}{\text{ln}(\Lambda/k_\text{ph})}$ is also shown (dashed).}
\label{figinteraction}
\end{figure}

The flow of $\lambda(k)$ for different initial values $\lambda_\Lambda$ is shown in Fig. \ref{figflowoflambda}. 
Following the flow from $\Lambda$ to $k_\text{ph}$ yields the dependence of $\lambda=\lambda(k_\text{ph})$ on $\lambda_\Lambda$ as displayed in Fig. \ref{figinteraction}.
It follows from Eq. \eqref{eqlambdakofk} that for positive initial values $\lambda_\Lambda$ the interaction strength $\lambda$ is bounded by
\begin{equation}
\lambda<\frac{4\pi}{\text{ln}(\Lambda/k_\text{ph})}\approx1.82.
\end{equation}
The last equation holds for our choice of $\Lambda$ and $k_\text{ph}$. We emphasize that our bound holds only if the interactions are approximately pointlike for all momenta below $\Lambda$. Close to a Feshbach resonance this may not be true and our formalism would need to be extended by considering nonlocal interactions or introducing an additional field for the exchanged two-atom state in the ``closed channel''. In other words, close to a Feshbach resonance the effective cutoff $\Lambda$ for the validity of a two-dimensional model with pointlike interactions may be substantially lower, and the upper bound on $\lambda$ correspondingly higher. 

\section{Flow equations at nonzero density}
In this section we investigate the many body problem for a nonvanishing density $n$ at zero temperature $T=0$. A crucial new ingredient as compared to the vacuum is the nonzero superfluid density 
\begin{equation}
n_S=\rho_0(k_\text{ph}).
\end{equation}
For interacting bosons at zero temperature the density $n$ and the superfluid density $n_S$ are equal. In contrast, the condensate density is given by the unrenormalized order parameter 
\begin{equation}
n_C=\bar{\rho}_0(k_\text{ph})=\bar{A}^{-1}(k_\text{ph})\rho_0(k_\text{ph}).
\end{equation}
Due to the repulsive interaction, $n_C$ may be smaller than $n$, the difference $n-n_C$ being the condensate depletion. (In the limit $\lambda\rightarrow 0$ we have $n=n_S=n_C$.) To obtain a nonzero density at temperature $T=0$ we have to go to positive chemical potential $\mu>0$. At the microscopic scale $k=\Lambda$ the minimum of the effective potential $U$ is then at $\rho_{0,\Lambda}=\mu/\lambda>0$. 

The superfluid density $\rho_0$ is connected to a nonvanishing ``renormalized order parameter'' $\phi_0$, with $\rho_0=\phi_0^*\phi_0$. It is responsible for an effective spontaneous breaking of the U(1)-symmetry. Indeed, the expectation value $\phi_0$ points out a direction in the complex plane so that the global U(1)-symmetry of phase rotations is broken by the ground state of the system. Goldstones theorem implies the presence of a gapless Goldstone mode, and the associated linear dispersion relation $\omega\sim|\vec{q}|$ accounts for superfluidity. The Goldstone physics is best described by using a real basis in field space by decomposing the complex field $\phi=\phi_0+\frac{1}{\sqrt{2}}(\phi_1+i\phi_2)$. Without loss of generality we can choose the expectation value $\phi_0$ to be real. The real fields $\phi_1$ and $\phi_2$ describe then the radial and Goldstone mode. respectively. For $\mu=\mu_0$ the inverse propagator reads in our truncation
\begin{equation}
G^{-1}=\bar{A}\begin{pmatrix} \vec{p}^2+V p_0^2+U^\prime+2\rho U^{\prime\prime}, & -S p_0 \\ S p_0, & \vec{p}^2+V q_0^2+U^\prime \end{pmatrix}.
\label{eqprop}
\end{equation}
Here $\vec{p}$ is the momentum of the collective excitation, and for $T=0$ the frequency obeys $\omega=-ip_0$.
In the regime with spontaneous symmetry breaking, $\rho_0(k)\neq 0$, the propagator for $\rho=\rho_0$ has $U^\prime=0$, $2\rho U^{\prime\prime}=2\lambda \rho_0\neq 0$, giving rise to the linear dispersion relation characteristic for superfluidity. This strongly modifies the flow equations as compared to the vacuum flow equations once $k^2 \ll 2\lambda \rho_0$. For $n\neq0$ the flow is typically in the regime with $\rho_0(k)\neq0$. In practice, we have to adapt the initial value $\rho_{0,\Lambda}$ such that the flow ends at a given density $\rho_0(k_\text{ph})=n$. For $k_\text{ph}\ll n^{1/2}$ one finds that $\rho_0(k_\text{ph})$ depends only very little on $k_\text{ph}$. As mentioned above, we will often choose the density to be unity such that effectively all length scales are measured in units of the interparticle distance $n^{-1/2}$. 

\begin{figure}
\includegraphics{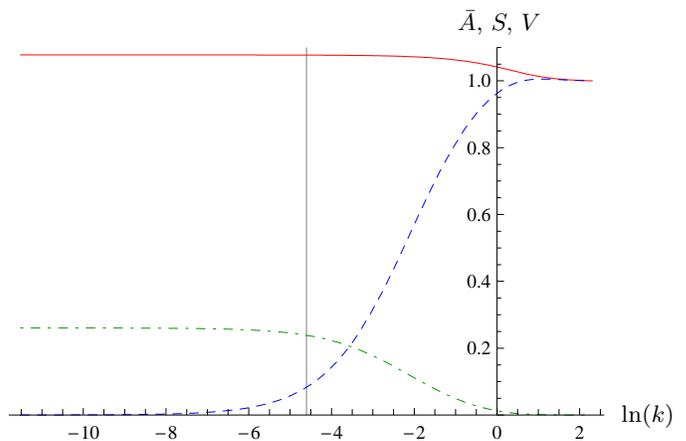}
\caption{(Color online) Flow of the kinetic coefficients $\bar{A}$ (solid), $S$ (dashed), and $V$ (dashed-dotted) at zero temperature $T=0$, density $n=1$, and vacuum interaction strength $\lambda=1$.}
\label{figFlowKinetic}
\end{figure}

In contrast to the vacuum with $T=\rho_0=0$, the flow of the propagator is nontrivial in the phase with $\rho_0>0$ and spontaneous U(1) symmetry breaking. In Fig. \ref{figFlowKinetic} we show the flow of the kinetic coefficients $\bar{A}$, $V$, $S$ for a renormalized or macroscopic interaction strength $\lambda=1$. 
The wavefunction renormalization $\bar{A}$ increases only a little at scales where $k\approx n^{1/2}$ and saturates then to a value $\bar{A}>1$. As will be explained below, we can directly infer the condensate depletion from the value of $\bar{A}$ at macroscopic scales. The coefficient of the linear $\tau$-derivative $S$ goes to zero for $k\rightarrow 0$. The frequency dependence is then governed by the quadratic $\tau$-derivative with coefficient $V$, which is generated by the flow and saturates to a finite value for $k\rightarrow 0$. 

\begin{figure}
\includegraphics{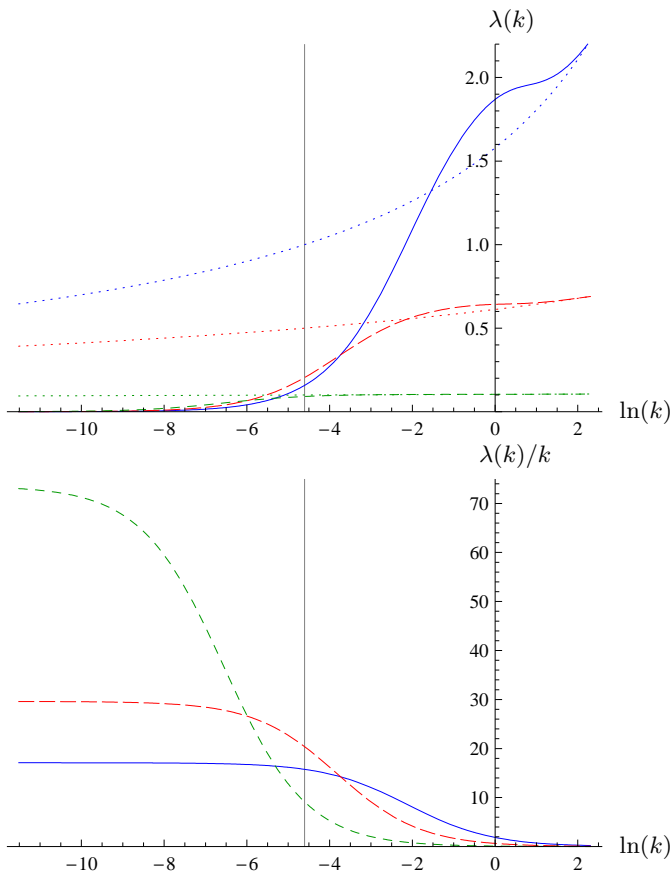}
\caption{(Color online) Flow of the interaction strength $\lambda(k)$ at zero temperature $T=0$, density $n=1$, for different initial values $\lambda_\Lambda$. The dotted lines are the corresponding graphs in the vacuum $n=0$. The vertical line labels our choice of $k_\text{ph}$. The lower plot shows $\lambda(k)/k$ for the same parameters, demonstrating that $\lambda(k)\sim k$ for small $k$.}
\label{figFlowLambda}
\end{figure}

The flow of the interaction strength $\lambda(k)$ for different values of $\lambda=\lambda(k_\text{ph})$ is shown in Fig. \ref{figFlowLambda}.
While the decrease with the scale $k$ is only logarithmic in vacuum, it becomes now linear $\lambda(k)\sim k$ for $k\ll n^{1/2}$. It is interesting that the ratio $\lambda(k)/k$ reaches larger values for smaller values of $\lambda_\Lambda$. 

\begin{figure}
\includegraphics{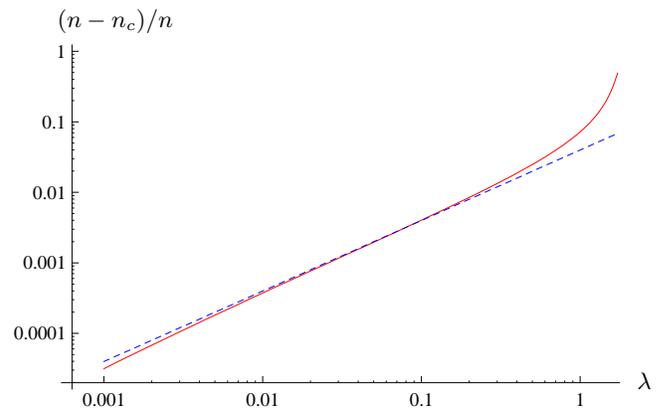}
\caption{(Color online) Condensate depletion $(n-n_c)/n$ as a function of the vacuum interaction strength $\lambda$. The dashed line is the Bogoliubov result $(n-n_c)/n=\frac{\lambda}{8\pi}$ for reference.}
\label{figDepletion}
\end{figure}

As $k$ is lowered from $\Lambda$ to $k_\text{ph}$, the renormalized order parameter or the superfluid density $\rho_0$ increases first and then saturates to $\rho_0=n=1$. This is expected since the superfluid density equals the total density at zero temperature. In contrast, the bare order parameter $\bar{\rho}_0=\rho_0/\bar{A}$ flows to a smaller value $\bar{\rho}_0<\rho_0$. As argued in \cite{FW}, the bare order parameter is just the condensate density, such that
\begin{equation}
n-n_C=\rho_0-\bar{\rho}_0=\rho_0(1-\frac{1}{\bar{A}})
\end{equation}
is the condensate depletion. Its dependence on the interaction strength $\lambda$ is shown in Fig. \ref{figDepletion}.
For small interaction strength $\lambda$ the condensate depletion follows roughly the Bogoliubov form
\begin{equation}
\frac{n-n_c}{n}=\frac{\lambda}{8\pi}.
\end{equation}
However, we find small deviations due to the running of $\lambda$ which is absent in Bogoliubov theory. For large interaction strength $\lambda\approx 1$ the deviation from the Bogoliubov result is quite substantial, since the running of $\lambda$ with the scale $k$ is more important. 

We also investigate the dispersion relation at zero temperature. The dispersion relation $\omega(p)$ follows from the condition
\begin{equation}
\text{det}\, G^{-1}(\omega(p),p)=0
\label{eqdispersionfromprop}
\end{equation}
where $G^{-1}$ is the inverse propagator after analytic continuation to real time $p_0\rightarrow i\omega$. As was shown in \cite{FW} the generation of the kinetic coefficient $V$ by the flow leads to the emergence of a second branch of solutions of Eq. \eqref{eqdispersionfromprop}. In our truncation the dispersion relation for the two branches $\omega_+(\vec{p})$ and $\omega_-(\vec{p})$ are
\begin{eqnarray}
\nonumber
\omega_\pm(\vec{p})&=&{\Bigg (}\frac{1}{V}(\vec{p}^2+\lambda \rho_0)+\frac{S^2}{2V^2}\\
\nonumber
&&\pm{\Bigg (}\left(\frac{1}{V}(\vec{p}^2+\lambda\rho_0)+\frac{S^2}{2V^2}\right)^2\\
&&-\frac{1}{V^2}\vec{p}^2(\vec{p}^2+2\lambda \rho_0){\Bigg )}^{1/2}{\Bigg )}^{1/2}.
\label{eqdispersionrelation}
\end{eqnarray}
In the limit $V\rightarrow 0$, $S\rightarrow 1$ we find that the lower branch approaches the Bogoliubov result $\omega_-\rightarrow \sqrt{\vec{p}^2(\vec{p}^2+2\lambda \rho_0)}$ while the upper branch diverges $\omega_+\rightarrow \infty$ and thus disappears from the spectrum. The lower branch is dominated by phase changes (Goldstone mode), while the upper branch reflects waves in the size of $\rho_0$ (radial mode). 

\begin{figure}
\includegraphics{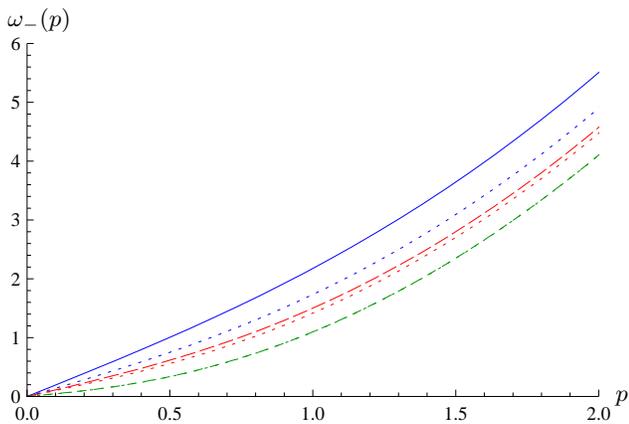}
\caption{(Color online) Lower branch of the dispersion relation $\omega_{-}(p)$ at temperature $T=0$ and for the vacuum interaction strength $\lambda=1$ (solid curve), $\lambda=0.5$ (upper dashed curve), and $\lambda=0.1$ (lower dashed curve). The units are set by the density $n=1$. We also show the Bogoliubov result for $\lambda=1$ (upper dotted curve) and $\lambda=0.5$ (lower dotted curve). For $\lambda=0.1$ the Bogoliubov result is identical to our result within the plot resolution.}
\label{figDispersionlinear}
\end{figure}

\begin{figure}
\includegraphics{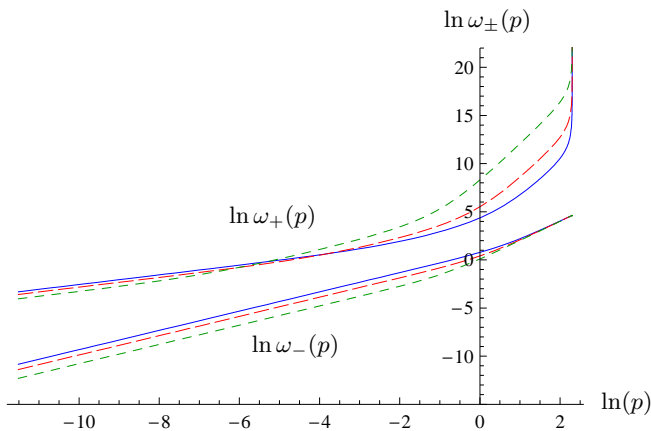}
\caption{(Color online) Dispersion relation $\omega_{-}(p)$, $\omega_{+}(p)$ at temperature $T=0$ and for vacuum interaction strength $\lambda=1$ (solid), $\lambda=0.5$ (long dashed), and $\lambda=0.1$ (short dashed). The units are set by the density $n=1$.}
\label{figDispersion}
\end{figure}

In principle, the coupling constants on the right-hand side of Eq. \eqref{eqdispersionrelation} also depend on the momentum $p=|\vec{p}|$. Since an external momentum provides an infrared cutoff of order $k\approx |\vec{p}|$ we can approximate the $|\vec{p}|$-dependence by using on the right-hand side of Eq. \eqref{eqdispersionrelation} the $k$-dependent couplings with the identification $k=|\vec{p}|$. Our result for the lower branch of the dispersion relation is shown in Fig. \ref{figDispersionlinear}.
We also plot the Bogoliubov result $\omega=\sqrt{\vec{p}^2(\vec{p}^2+2\lambda \rho_0)}$ for comparison. For small $\lambda$ our result is in agreement with the Bogoliubov result, while we find substantial deviations for large $\lambda$. Both branches $\omega_+$ and $\omega_-$ are shown in Fig. \ref{figDispersion} on a logarithmic scale. 
Since we start with $V=0$ at the microscopic scale $\Lambda$ we find $\omega_+(\vec{p})\rightarrow\infty$ for $|\vec{p}|\rightarrow \Lambda$.

\begin{figure}
\includegraphics{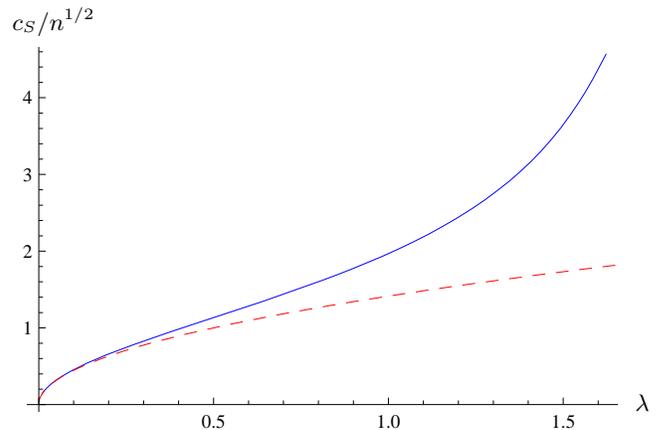}
\caption{(Color online) Dimensionless sound velocity $c_S/n^{1/2}$ as a function of the vacuum interaction strength (solid). We also show the Bogoliubov result $c_S=\sqrt{2\lambda \rho_0}$ for reference (dashed).}
\label{figsoundvelocity}
\end{figure}

The sound velocity $c_S$ can be extracted from the dispersion relation. More precisely, we compute the microscopic sound velocity for the lower branch $\omega_-(\vec{p})$ as $c_S=\frac{\partial \omega}{\partial p}$ at $p=0$. In our truncation we find \cite{FW}
\begin{equation}
c_S^2=\frac{2\lambda\rho_0}{S^2+2\lambda\rho_0 V}.
\end{equation}
Our result for $c_S$ at $T=0$ is shown in Fig. \ref{figsoundvelocity} as a function of the interaction strength $\lambda$. 
For a large range of small $\lambda$ we find good agreement with the Bogoliubov result $c_S^2=2\lambda \rho_0$. However, for large $\lambda$ or result for $c_S$ exceeds the Bogoliubov result by a factor up to 2.

\section{Nonzero Temperature}
At nonzero temperature and for infinite volume, long range order is forbidden in two spatial dimensions by the Mermin-Wagner theorem. Because of that, no proper Bose-Einstein condensation is possible in a two-dimensional homogeneous Bose gas at nonvanishing temperature. However, even if the order parameter vanishes in the thermodynamic limit of infinite volume, one still finds a nonzero superfluid density for low enough temperature. The superfluid density can be considered as the square of a renormalized order parameter $\rho_0=|\phi_0|^2$ and the particular features of the low-temperature phase can be well understood by the physics of the Goldstone boson for a phase with effective spontaneous symmetry breaking \cite{Wetterich:1991be}. The renormalized order parameter $\phi_0$ is related to the expectation value of the bosonic field $\bar{\phi}_0$ and therefore to the condensate density $\bar{\rho}_0=\bar{\phi}_0^2$ by a wave function renormalization, defined by the behavior of the bare propagator $\bar{G}$ at zero frequency for vanishing momentum
\begin{equation}
\phi_0=\bar{A}^{1/2}\bar{\phi}_0,\quad \rho_0=\bar{A}\bar{\rho}_0,\quad \bar{G}^{-1}(\vec{p}\rightarrow 0)=\bar{A}\vec{p}^2.
\end{equation}
While the renormalized order parameter $\rho_0(k)$ remains nonzero for $k\rightarrow 0$ if $T<T_c$, the condensate density $\bar{\rho}_0=\rho_0/\bar{A}$ vanishes since $\bar{A}$
diverges with the anomalous dimension, $\bar{A}\sim k^{-\eta}$. After restoring dimensions the relation
\begin{equation}
\rho_0=\lim\limits_{\vec{p}\to 0}{\frac{\bar{\rho}_0}{\vec{p}^2\bar{G}(\vec{p})}}
\end{equation}
is the Josephson relation \cite{Josephson}. 

The strict distinction between a zero Bose-Einstein condensate $\bar{\rho}_0=0$ and a nonzero superfluid density $\rho_0>0$ for nonzero temperature $0<T<T_c$ is valid only in the infinite volume limit of a homogeneous system. For a finite size of the system, as atoms in a trap, the running of $\bar{A}(k)$ is effectively stopped at some scale $k_\text{ph}$. There are simply no collective modes with wavelength larger than the size of the system, whose fluctuations would be responsible for a further increase of $\bar{A}$. With a finite $\bar{A}_\text{ph}$ both $\bar{\rho}_0$ and $\rho_0$ are nonzero for $T<T_c$, and the distinction between a Bose-Einstein condensate and superfluidity is no longer relevant in practice. For large systems $\bar{A}(k_\text{ph})$ can be large, however, such that the condensate density can be suppressed substantially as compared to the superfluid density. In any case, there is only one critical temperature $T_c$, defined by $\rho_0(T<T_c)>0$.

\begin{figure}
\includegraphics{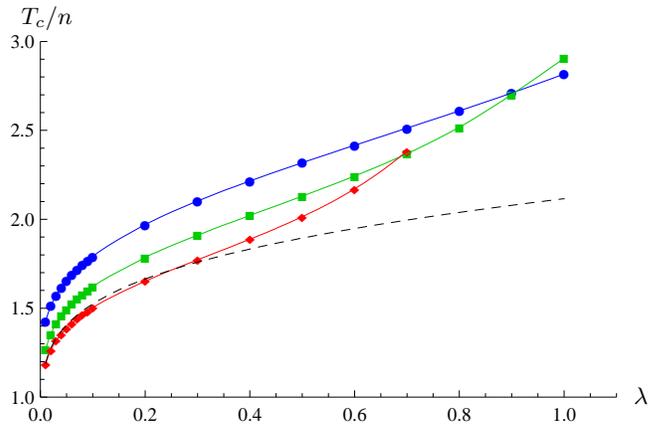}
\caption{(Color online) Critical temperature $T_c/n$ as a function of the interaction strength $\lambda$. We choose here $k_\text{ph}=10^{-2}$ (circles), $k_\text{ph}=10^{-4}$ (boxes) and $k_\text{ph}=10^{-6}$ (diamonds). For the last case the bound on the scattering length is $\lambda<\frac{4\pi}{\text{ln}(\Lambda/k_\text{ph})}\approx 0.78$. We also show the curve $\frac{T_c}{n}=\frac{4\pi}{\text{ln}(\zeta/\lambda)}$ (dashed) with the Monte-Carlo result $\zeta=380$ \cite{PRS} for reference.}
\label{figtcoflambda}
\end{figure}

The flow equations permit a straightforward computation of $\rho_0(T)$ for arbitrary $T$, once the interaction strength of the system has been fixed at zero temperature and density. We have extracted the critical temperature as a function of $\lambda=\lambda(k_\text{ph})$ for different values of $k_\text{ph}$. The behavior for small $\lambda$,
\begin{equation}
\frac{T_c}{n}=\frac{4\pi}{\text{ln}(\zeta/\lambda)}
\label{Tcperturbative}
\end{equation}
is compatible with the free theory where $T_c$ vanishes for $k_\text{ph}\rightarrow0$ and with the perturbative analysis in Ref. \cite{d2BosePerturbative}. We find that the value of $\zeta$ depends on the choice of $k_\text{ph}$. For $k_\text{ph}=10^{-2}$ we find $\zeta=100$, while $k_\text{ph}=10^{-4}$ corresponds to $\zeta=225$ and $k_\text{ph}=10^{-6}$ to $\zeta=424$. In Fig. \ref{figtcoflambda} we show or result for $T_c/n$ as a function of $\lambda$ for these choices. We also plot the curve in Eq. \eqref{Tcperturbative} with the Monte-Carlo result $\zeta=380$ from Ref. \cite{PRS}.
We find that $T_c$ vanishes for $k_\text{ph}\to0$ in the interacting theory as well. This is due to the increase of $\zeta$ and, for a fixed microscopical interaction, to the decrease of $\lambda(k_\text{ph})$. Since the vanishing of $T_c/n$ is only logarithmic in $k_\text{ph}$, a phase transition can be observed in practice. We find agreement with Monte-Carlo results \cite{PRS} for small $\lambda$ if $k_\text{ph}/\Lambda\approx 10^{-7}$. The dependence of $T_c/n$ on the size of the system $k_\text{ph}^{-1}$ remains to be established for the Monte-Carlo computations.

The critical behavior of the system is governed by a Kosterlitz-Thouless phase transition. Usually this is described by considering the thermodynamics of vortices. In Refs. \cite{Grater:1994qx,VonGersdorff:2000kp} it was shown that functional renormalization group can account for this ``nonperturbative'' physics without explicitly taking vortices into account. The correlation length in the low-temperature phase is infinite. In our picture, this arises due to the presence of a Goldstone mode if $\rho_0>0$. The system is superfluid for $T<T_c$. The powerlike decay of the correlation function at zero frequency 
\begin{equation}
\bar{G}(\vec{p})\sim (\vec{p}^2)^{-1+\eta/2}
\label{eq:momentumdependenceofpropagator}
\end{equation}
is directly related to the running of $\bar{A}$. As long as $k^2\gg \vec{p}^2$ the bare propagator obeys approximately
\begin{equation}
\bar{G}=\frac{1}{\bar{A}(k)\vec{p}^2},\quad \bar{A}(k)\sim k^{-\eta}.
\label{eq:momentumdependenceofpropfinitek}
\end{equation}
Once $k^2\ll \vec{p}^2$, the effective infrared cutoff is given by $\vec{p}^2$ instead of $k^2$, and therefore $\bar{A}(k)$ gets replaced by $\bar{A}(\sqrt{\vec{p}^2})$, turning Eq. \eqref{eq:momentumdependenceofpropfinitek} into Eq. \eqref{eq:momentumdependenceofpropagator}. For large $\rho_0$ the anomalous dimension depends on $\rho_0$ and $T$, $\eta=T/(4\pi \rho_0)$.

Another characteristic feature of the Kosterlitz-Thouless phase transition is a jump in the superfluid density at the critical temperature. However, a true discontinuity arises only in the thermodynamic limit of infinite volume ($k_\text{ph}\rightarrow 0$), while for finite systems ($k_\text{ph}>0$) the transition is smoothened. In order to see the jump, as well as essential scaling for $T$ approaching $T_c$ from above, our truncation is insufficient. These features become visible only in extended truncations that we will briefly describe next. 

For very small scales $\frac{k^2}{T}\ll 1$, the contribution of Matsubara modes with frequency $q_0=2\pi T n$, $n\neq 0$, is suppressed since nonzero Matsubara frequencies act as an infrared cutoff. In this limit a dimensionally reduced theory becomes valid. The long distance physics is dominated by classical two-dimensional statistics, and the time dimension parametrized by $\tau$ no longer plays a role. 

The flow equations simplify considerably if only the zero Matsubara frequency is included, and one can use more involved truncations. Such an improved truncation is indeed needed to account for the jump in the superfluid density. In Ref. \cite{VonGersdorff:2000kp} the next to leading order in a systematic derivative expansion was investigated. It was found that for $k\ll T$ the flow equation for $\rho_0$ can be well approximated by
\begin{equation}
\partial_t \rho_0=2.54\,T^{-1/2}(0.248\, T-\rho_0)^{3/2}\,\theta(0.248 \,T-\rho_0).
\label{eq:improvedflowofrho}
\end{equation}
We switch from the flow equation in our more simple truncation to the improved flow equation \eqref{eq:improvedflowofrho} for scales $k$ with $k^2/T<10^{-3}$. We keep all other flow equations unchanged. A similar procedure was also used in Ref. \cite{DrHCK}.

\begin{figure}
\includegraphics{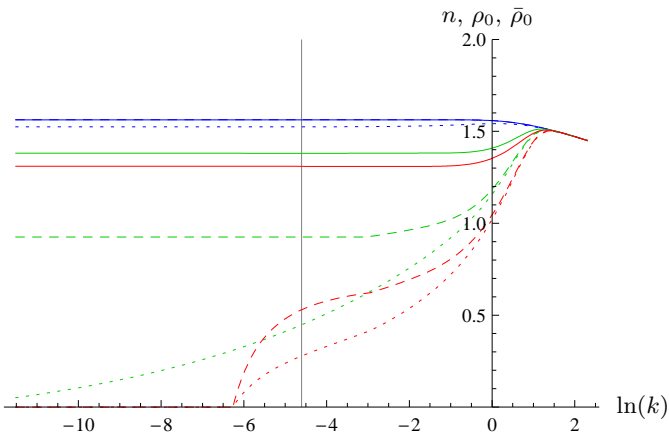}
\caption{(Color online) Flow of the density $n$ (solid), the superfluid density $\rho_0$ (dashed), and the condensate density $\bar{\rho}_0$ (dotted) for chemical potential $\mu=1$, vacuum interaction strength $\lambda=0.5$ and temperatures $T=0$ (top), $T=2.4$ (middle) and $T=2.8$ (bottom). The vertical line marks our choice of $k_\text{ph}$. We recall $n=\rho_0$ for $T=0$ such that the upper dashed and solid lines coincide.}
\label{figFlowofnrho}
\end{figure}

\begin{figure}
\includegraphics{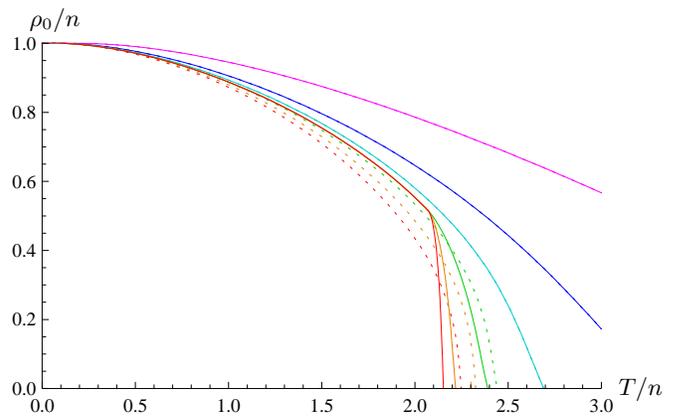}
\caption{(Color online) Superfluid fraction of the density $\rho_0/n$ as a function of the dimensionless temperature $T/n$ for interaction strength $\lambda=0.5$ at different macroscopic scales $k_\text{ph}=1$ (upper curve), $k_\text{ph}=10^{-0.5}$, $k_\text{ph}=10^{-1}$, $k_\text{ph}=10^{-1.5}$, $k_\text{ph}=10^{-2}$, $k_\text{ph}=10^{-2.5}$ (bottom curve). We plot the result obtained with the improved truncation for small scales (solid) as well as the result obtained with our more simple truncation (dotted). (The truncations differ only for the three lowest lines.)}
\label{figrhodnoftemperature}
\end{figure}

\begin{figure}
\includegraphics{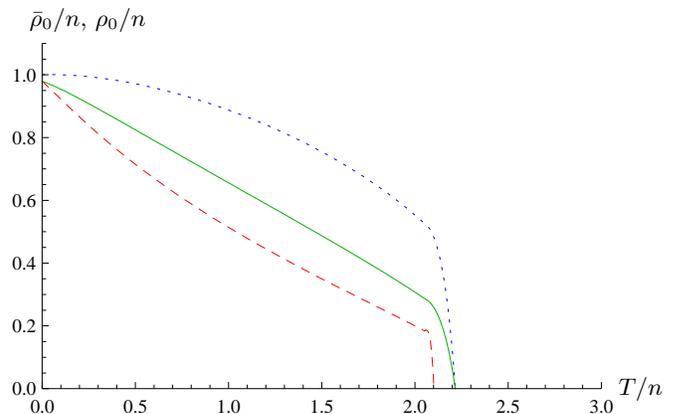}
\caption{(Color online) Condensate fraction of the density $\bar{\rho}_0/n$ as a function of the dimensionless temperature $T/n$ for interaction strength $\lambda=0.5$ at macroscopic scale $k_\text{ph}=10^{-2}$ (solid curve) and $k_\text{ph}=10^{-4}$ (dashed curve). For comparison, we also plot the superfluid density $\rho_0/n$ at $k_\text{ph}=10^{-2}$ (dotted). These results are obtained with the improved truncation.}
\label{figcondensatedensity}
\end{figure}

In Fig. \ref{figFlowofnrho} we show the flow of the density $n$, the superfluid density $\rho_0$ and the condensate density $\bar{\rho}_0$ for different temperatures.
In Fig. \ref{figrhodnoftemperature} we plot our result for the superfluid fraction of the density as a function of the temperature for different scales $k_\text{ph}$. 
One can see that with the improved truncation the jump in the superfluid density is indeed found in the limit $k_\text{ph}\rightarrow 0$. Fig. \ref{figcondensatedensity} 
shows the condensate fraction $\bar{\rho}_0/n$ and the superfluid density fraction $\rho_0/n$ as a function of $T/n$. We observe the substantial $k_\text{ph}$ dependence of the condensate fraction, as well as an effective jump at $T_c$ for small $k_\text{ph}$. We recall that the infinite volume limit $k_\text{ph}=0$ amounts to $\bar{\rho}_0=0$ for $T>0$. 

The Kosterlitz-Thouless description is only valid if the zero Matsubara frequency mode ($n=0$) dominates. For a given nonzero $T$ this is always the case if the the characteristic length scale goes to infinity. In the infinite volume limit the characteristic length scale is given by the correlation length $\xi$. The description in terms of a classical two dimensional system with U(1) symmetry is the key ingredient of the Kosterlitz-Thouless description and holds for $\xi^2 T\gg 1$. In the infinite volume limit this always holds for $T<T_c$ or near the phase transition, where $\xi$ diverges or is very large. For a finite size system the relevant length scale becomes $k_\text{ph}^{-1}$ if this is smaller than $\xi$. Thus the Kosterlitz-Thouless picture holds only for $T>k_\text{ph}^2$. 

For very small temperatures $T<k_\text{ph}^2$ one expects a crossover to the characteristic behavior near a quantum critical phase transition, governed by the quantum critical fixed point. The crossover between the different characteristic behaviors for $T>k_\text{ph}^2$ and $T<k_\text{ph}^2$ can be observed in several quantities. As an example we may take Fig. \ref{figFlowofnrho} and compare the flow of $\rho_0$ and $\bar{\rho}_0$ for low $T$ (close to the $T=0$ curve) or large $T$ (other curves). 

\section{Conclusions}
\label{sectConclusions}
In conclusion, we have demonstrated that functional renormalization for the average action yields a unified picture for ultracold nonrelativistic bosons in arbitrary dimension $d$. We have employed in this paper the same simple truncation for $d=2$ as in Ref. \cite{FW} for $d=3$. While for $d=3$ the perturbative Bogoliubov treatment is reliable for a wide range of parameters, we find for $d=2$ substantial deviations if the coupling $\lambda$ is large, or for the regime close to the phase transition. The Kosterlitz-Thouless type of the transition is nonperturbative in $\lambda$ and cannot be seen within the Bogoliubov approximation. 

In particular, we have obtained the following main results for interacting Bose gases in two dimensions:
\begin{enumerate}
	\item An upper bound on the interaction strength $\lambda$ in vacuum arises if the characteristic momentum of the scattering particle $p=|\vec{p}|$ is sufficiently below the ultraviolet scale $\Lambda$ below which a two-dimensional description with pointlike interactions becomes valid,
\begin{equation}
\lambda(p)<\frac{4\pi}{\text{ln}(\Lambda/p)}.
\end{equation}
The dependence of $\lambda(p)$ on $p$ is logarithmic, $\partial \lambda /\partial \text{ln} p=\lambda^2/(4\pi)$.
	\item The condensate depletion at $T=0$ is reasonably well approximated by the Bogoliubov result if $\lambda$ is small, $(n-n_C)/n=\lambda/(8\pi)$. Logarithmic corrections arise due to the running of $\lambda$. For large $\lambda$ the repulsive interaction suppresses the condensate and the condensate depletion can exceed the Bogoliubov result by a large factor.
	\item For the dispersion relation at $T=0$ we observe a substantial deviation from the Bogoliubov result $\omega=p\sqrt{p^2+2\lambda n}$ for large $\lambda$. We also find a second branch of ``radial excitations''. The lower branch is dominated by phase fluctuations and shows the characteristic superfluid behavior $\omega\sim p$ for $p\to 0$. The upper branch corresponds to excitations of the local superfluid density $\rho_0$.
	\item The sound velocity at $T=0$ exceeds the Bogoliubov result for large $\lambda$ by up to a factor $2$, while the perturbative result is valid for small $\lambda$. 
	\item The system shows a second order phase transition. The critical temperature $T_c$ depends logarithmically on $\lambda$, $T_c/n=4\pi/\text{ln}(\zeta/\lambda)$. The coefficient $\zeta$ depends logarithmically on the size of the system $l=k_\text{ph}^{-1}$. For $k_\text{ph}\approx 10^{-7} \Lambda$ we find $\zeta\approx 400$ and agreement with Monte-Carlo simulations. The $l$-dependence of $T_c/n$ still needs to be established by numerical simulations.
	\item The phase transition is found to be of the Kosterlitz-Thouless type, with a jump of the superfluid density $\Delta \rho_0\approx T_c/4$. For a finite size system this jump is smoothened. Below $T_c$, the correlation length is infinite. The correlation function shows a powerlike decay, with an anomalous dimension $\eta$, $\bar G(p)\sim p^{-2+\eta}$. The anomalous dimension depends on the superfluid density $\rho_0$ and on the temperature $\eta=T/(4\pi \rho_0)$.
	\item We have computed the superfluid density $\rho_0$ and the condensate density $n_C=\bar{\rho}_0$ as a function of $T$. The condensate density vanishes in the infinite volume limit $l\to \infty$, if $T>0$. For finite $l$ a nonvanishing condensate can be observed.
\end{enumerate}

In this paper we have employed a rather simple truncation which is sufficient to show all the characteristic features of two-dimensional cold and dilute nonrelativistic bosons. One exception is the jump in the superfluid density which needs an improved truncation in the range of very small $k^2/T$. Extensions of the truncation are straightforward and should improve the quantitative accuracy for large $\lambda$. With increasing experimental precision the verification of a quantitatively precise theoretical computation will be a challenge.

\end{document}